\documentclass[12pt,reqno]{article}
\usepackage{amsmath,amsthm,amssymb,epsfig,alltt}
\usepackage{setspace}

\newcommand{\nn}{\nonumber }

\def\a{\alpha}

\def\e{\epsilon}
\def\p{\partial}
\def\m{\mu}
\def\n{\nu}
\def\t{\tau}
\def\th{\theta}
\def\s{\sigma}
\def\g{\gamma}

\def\half{\frac{1}{2}}

\def\barz{{\bar z}}

\def\sp{\sigma^\prime}
\def\nn{\nonumber}

\def\2pap{2\pi\alpha^\prime}

\def\beq{\begin{eqnarray}}
 \def\eeq{\end{eqnarray}}
 \def\4pap{4\pi\a^\prime}
 
 \def\sp{{\s^\prime}}
 
 \def\ap{{\a^\prime}}

 \def\barz{{\bar z}}

\begin{document}


\title{Applications of Thirring Model to Inhomogenous Rolling Tachyon and Dissipative Quantum Mechanics}

\author{Taejin Lee
\\~~\\
Department of Physics, Kangwon National University \\
Chuncheon 200-701, Korea 
\footnote{Email: taejin@kangwon.ac.kr,~~~ 
}
\\ ~~ \\
Pacific
Institute for Theoretical Physics Department of Physics and \\
Astronomy, University of British Columbia \\ 6224 Agricultural
Road, Vancouver, British Columbia V6T 1Z1, Canada}

\maketitle

\centerline{\bf \large Abstract}~~\\
We study the rolling tachyon and the dissipative quantum mechanics using the Thirring model with a boundary mass. 
We construct a boundary state for the dissipative quantum system in one dimension, which describes the system at the off-critical points as well as at the critical point. Then we extend the Thirring model with a boundary mass in order to depict the time evolution of an unstable D-branes with one direction wrapped on a circle of radius $R$, which is termed the inhomogeneous rolling tachyon. The analysis based on the Thirring model shows that the time dependent evolution of the inhomogeneous tachyon is possible only when $\frac{2}{\sqrt{3}}< R < 2$.
\\



\newpage


\section{Introduction}

The two space-time dimensional quantum field theory of massless
bosons with a periodic boundary potential has recently come into the
spotlight again. This boundary conformal field theory has received 
constant attention along the years, since it describes 
the various condensed physics systems such as
the dissipative quantum mechanics of a particle in a one-dimensional periodic potential \cite{schmid,Guinea,fisher,furusaki,Callan:1989mm}, 
Josephson junction arrays \cite{larkin,fazio,sodano} and the
dissipative Hofstadter problem \cite{Callan:1991da,Callan:1992vy}.
The applications of the theory also include the Kondo problem
\cite{Affleck:1990iv,Affleck:1990by}, the one-dimensional
conductors \cite{Kane:1992}, tunnelling between Hall edge states
\cite{kane2}, and junctions of quantum wires~\cite{Oshikawa:2005fh}.
The recent revival of interest in the theory is mainly due to the
string theory applications. As the string with its ends on an unstable D-brane develops a marginal boundary interaction, the tachyon field of the open string condensates \cite{Witten:1992qy,Shatashvili:1993kk,Shatashvili:1993ps,Tlee:01nc,Tlee:01os,Tlee:01one}. This process, called ``rolling tachyon" \cite{Sen:2002nu,senreview,lambert,sen022,mukhopa,sen023,sen0305,sen031,sen0306,sen0308,sen04,larsen02,gibbons,okuda,kim,hlee,rey,rey2,taka,doug,
arefeva,demasure,gutperle03,larsen,
constable,schomerus,nagami,foto,coletti,forini,
Tlee:06,gianluca}, is believed to be responsible for the decay of the unstable D-brane. 

Recently, the theory with a periodic boundary potential is discussed in detail
by applying the fermionization technique \cite{Tlee:2005ge,Hassel}. The advantage of the fermion formulation is that one can explicitly construct the boundary state for the rolling tachyon, since the periodic boundary potential becomes a boundary fermion mass term, which is quadratic in fermion field.
The fermion formulation of the theory, however, has been performed only for the theory at the critical point. In this paper we apply the fermion formulation to the theory at off-critical points to develop a perturbation theory around the critical points. In the fermion formulation, the theory at off-critical points is described by the Thirring model with a boundary mass. 
The fermion formulation of the theory at off-critical points would be 
also useful to discuss the inhomogeneous rolling tachyon.

The massless Thirring model \cite{Thirring}, which is the first example of an exactly solvable relativistic interacting field theory, has served as an excellent laboratory for the study of various aspects of the quantum field theory in two dimensions. Since the seminal work of Thirring, extensive studies of the model have been carried out by numerous authors \cite{Glaser,Johnson,Hagen,Klaiber}; notably
the complete solution of the theory was obtained by Klaiber \cite{Klaiber}.
Along this line, Schwinger \cite{Schwinger} obtained an exact solution of quantum electrodynamics in 1+1 dimensions and Coleman \cite{Coleman} proved the equivalence between the sine-Gordon model and the massive Thirring model.
 
Here we widen the range of applications of the Thirring model by studying the dissipative quantum system and the inhomogeneous rolling tachyon in terms of it. The Thirring model with a boundary mass is found to be the most suitable framework to discuss the dissipative quantum system: At the critical point the Thirring coupling,
which is directly related to the friction constant of the system,
vanishes and the theory reduces to a free fermion theory with a boundary mass. The boundary state takes a simple form at the critical point when it is expressed in fermion variables
as discussed in refs. \cite{Tlee:2005ge,Hassel,ji}. Thus, the Thirring model provides a perturbation theory expanded in the Thirring coupling near the critical point. 
One of advantages of the boundary state formulation is that all perturbative corrections can be easily shown to vanish at the critical point.



Recently, Sen \cite{sen022}
discussed a $Dp$-brane of the bosonic string theory with one direction wrapped on a cricle of radius $R >1$. Since the boundary interaction depends on the additional spatial compact coordinate $Y$, we expect that the tachyon condensation may be inhomogeneous. In this paper we show that the inhomogeneous rolling tachyon can be also studied within the framework of the Thirrring model and the boundary state formulation.

\section{The Schmid Model and The Thirring Model}

Caldeira and Leggett \cite{caldeira83ann,caldeira83phy} discussed first
the quantum mechanical description of dissipation by introducing a bath or environment, which consists of an infinite number of harmonic oscillators, coupled to the system. 
In the quantum theory the interaction with the bath produces a non-local effective interaction. Subsequenlty Schmid \cite{schmid} studied the dissipative system in the presence of a periodic potential. The one dimensional dissipative model with a periodic potential, called Schmid model, is described by the following action
\beq 
S_{SM} &=& \frac{\eta}{4\pi \hbar} \int^{T/2}_{-T/2} dt dt^\prime 
\frac{\left(X(t) - X(t^\prime)\right)^2}{(t-t^\prime)^2} 
- \frac{V_0}{\hbar} \int^{T/2}_{-T/2} dt \cos \frac{2\pi X}{a}. 
\eeq
The first non-local term is responsible for the dissipation, 
and the second term denotes the periodic potential respectively. An interesting feature of the model is that it exhibits a phase transition, unlike one dimensional quantum mechanical systems with local interactions only. Depending on the value of the friction constant $\eta$, the phase diagram of the system divides into two phases; the localized phase and the delocalized one. 

We can map the Schmid model to the string theory on a disk
by identifying the time as the boundary parameter $\sigma$ 
in string theory and scaling the field variable $X$ as follows:
\beq
t = \frac{T}{2\pi} \tau, \quad X \rightarrow \frac{a}{2\pi} X.
\eeq
Then, the action for the Schmid model reads as 
\beq
S_{SM} &=& \frac{\eta}{4\pi \hbar} \left(\frac{a}{2\pi}\right)^2 
\int d\tau d\tau^\prime \,
\frac{\left(X(\t) - X(\t^\prime)\right)^2}{(\t-\t^\prime)^2} 
- \frac{V_0}{\hbar} \frac{T}{2\pi} \int d\tau \,\half
\left(e^{iX} + e^{-iX} \right). 
\eeq
This action can be interpreted as the boundary effective action 
for the open bosonic string subject to a boundary periodic potential on a disk with a boundary condition; on the boundary
$\p M$, $X(\s,\t) = X(\t)$,
\beq
e^{-iS_{SM}} &=& \int D[X] \exp\Biggl[ - i\left(\frac{1}{4\pi \ap}
\int_M d\t d\s \p_\a X \p^\a X  - \frac{m}{2}
\int_{\p M} d\t \left(e^{iX}+ e^{-iX} \right)\right)\Biggr]. 
\eeq 
Here, we identify the physical parameters of the two theories as 
\beq
\frac{\eta}{4\pi \hbar} \left(\frac{a}{2\pi}\right)^2 
= \frac{1}{8\pi^2 \ap}, \quad 
-\frac{V_0}{\hbar} \frac{T}{2\pi} = m.
\eeq
In string theory the periodic potential describes the interaction
between the open string the unstable D-brane. 

The open string dynamics is often described more efficiently in its equivalent closed string picture by the boundary state formulation. The corresponding closed string action is obtained from its open string action by simply taking $\s \rightarrow \t$, $\t \rightarrow \s$,
\beq \label{sine1}
S = \frac{1}{4\pi \ap}
\int d\t d\s\, \p_\a X \p^\a X  - \frac{m}{2}
\int d\s \left(e^{iX}+ e^{-iX} \right).
\eeq
The open string dynamics can be encoded completely by the boundary state.

When $\a= 1/\ap= 1$, the system becomes critical. This can be
easily understood if we introduce an auxiliary boson field $Y$
and fermionize the system \cite{Tlee:2005ge,Hassel,Polchinski:1994my}. 
Introducing an auxiliary boson field
$Y$ which satisfies the Dirichlet condition $Y|_{\p M} = 0$ at the
boundary, and defining the boson fields,
$\phi_1 = \frac{X+Y}{\sqrt{2}}, ~~~
\phi_2 = \frac{X-Y}{\sqrt{2}}$, 
we may rewrite the action as
\beq \label{sine2}
S =  \frac{\a}{4\pi}\int_M d\t d\s \sum_i^2 \p \phi_i \p \phi_i
- \frac{m}{4} \int_{\p M} d\t \sum_i^2 \left(e^{i\sqrt{2}\phi_i}+ e^{-i\sqrt{2} \phi_i}\right).
\eeq
Since $Y$ is a free boson field and it vanishes at the boundary,
$Y$ is completely decoupled from the physical degrees of freedom. 
We see that if $\a=1$, 
$e^{\pm i\sqrt{2} \phi_i}$ are marginal boundary operators with the scaling dimension $1$. 
An explicit calculation of the current correlation function or the mobility shows that the theory becomes indeed critical where $\a =1$
\beq
\langle 0| \p_\s X(\s) \p_\s X(\sp) |B \rangle = - \frac{1}{2} (1-\pi^2 m^2) 
\sin^{-2}\frac{(\s-\sp)}{2}.
\eeq

The boundary state at the critical point can be explicitly evaluated
if the model is fermionized; the boson fields are mapped to the fermion fields as 
\begin{subequations}
\label{generallabel}
\begin{eqnarray}\label{fermionization1}
\psi_{1L}(z)&=&\zeta_{1L}:e^{-\sqrt{2}i\phi_{1L}(z)}:,~~
\psi_{2L}(z)=\zeta_{2L}:e^{\sqrt{2}i\phi_{2L}(z)}: \\
\psi_{1R}(\bar z)&=&\zeta_{1R}:e^{\sqrt{2}i\phi_{1R}(\bar z)}:,~~
\psi_{2R}(\bar z) = \zeta_{2R} :e^{-\sqrt{2}i\phi_{2R}(\bar z)}:
\label{fermionization2}
\end{eqnarray}
\end{subequations}
where $\zeta_{iL/R}$ are co-cycles, ensuring the anti-commutation
relations between the fermion operators. Since the boundary interaction term can be written as a boundary fermion mass term, which is only quadratic in fermion field, the model is exactly solvable
\beq
S &=& \int \frac{d\tau d\sigma}{2\pi} ~
\left(\bar\psi_1 \g^\m \p_\m \psi_1 + \bar\psi_2 \g^\m \p_\m \psi_2\right) + m \int \frac{d\s}{2\pi} \left(\bar\psi_1 \psi_1
+ \bar\psi_2 \psi_2 \right)
\eeq
where $\psi_i = (\psi_{iL}, \psi_{iR})^t$, and
\beq
\gamma^0 &=& \s_1, \quad
\gamma^1 = \s_2, \quad 
\gamma^5 = \s_3 = -i \gamma^0 \gamma^1. 
\eeq
The boundary state is given formally as
\beq
|B \rangle = :\exp\left[m \int \frac{d\s}{2\pi} \left(\bar\psi_1 \psi_1 + \bar\psi_2 \psi_2 \right)\right]:|N,D\rangle
\eeq
where $|N,D\rangle$ is a simple boundary state satisfying
\beq \label{simple}
\left(\psi_R(0,\s) +i \s^2 \psi_L(0,\s)\right)|N,D\rangle = 0,~~
\left(\psi^\dagger_R(0,\s) +i \psi^\dagger_L(0,\s)\s^2\right)|N,D\rangle =0.
\eeq
We refer the reader to ref. \cite{Hassel} for the explicit expression of the boundary state $|B\rangle$. 

Now let us discuss the dissipative system off the critical points. When $\a \not =1$, we 
may write the action as
\beq
S &=&  \frac{1}{4\pi}\int_M d\t d\s \sum_i^2 \p \phi_i \p \phi_i
+ \frac{1}{4\pi}\left(\a -1 \right)\int_M d\t d\s \sum_i^2 \p \phi_i \p \phi_i \nn\\
&& - \frac{m}{4} \int_{\p M} d\t \sum_i^2 \left(e^{i\sqrt{2}\phi_i}+ e^{-i\sqrt{2} \phi_i}\right).
\eeq
and treat the second term as an interaction. In terms of the fermion fields the second term can be written as the Thirring interaction term.
Hence, the fermionized action is given by
\beq
S &=&  \frac{1}{2\pi} \int_M d\t d\s \sum_i^2
\left(\bar{\psi}_i \gamma^\m \p_\m 
\psi_i + \frac{g}{4\pi} j^\m_i j_{i\m}\right) 
+ \frac{m}{2} \int_{\p M} d\s \sum_i^2 {\bar \psi}_i \psi_i
\eeq
where $g = \pi (\a -1)$.
This is the Thirring model with a boundary mass. At the critical point where $g =0$ ($\a = 1$), the action reduces to the free fermion theory with a boundary mass. Near the critical point, we can use this Thirring action to develop perturbation
theory for the dissipative quantum system. 
We note that the theory at the critical point corresponds to the 
homogeneous rolling tachyon of the string theory \cite{Sen:2002nu}.


\section{Boundary State near the Critical Point}

In order to apply the boundary state formulation to the
dissipative system near the critical point, the bulk action 
should be free. We may transmute the bulk Thirring interaction
into a boundary one by introducing Abelian gauge fields
\beq
S &=& \frac{1}{2\pi} \int_M d\t d\s\sum_i^2
\left[
\bar{\psi}_i\gamma^\m\left(\p_\m + iA_{i\m}\right)\psi_i 
+ \frac{\pi}{g} A_{i\m} A_i{}^\m \right] 
+  \frac{m}{2} \int_{\p M} d\s \sum_i^2 {\bar \psi}_i \psi_i .
\eeq
Since in general the Abelian gauge vector fields in $1+1$ dimensions may be decomposed as 
\beq
A^\m_i = \e^{\m\n} \p_\n \th_i + \p^\m \chi_i, ~~~ i =1,2 ,
\eeq
the interaction between the gauge fields and the fermion fields
may be removed by a gauge transformation
\beq
\psi_i = e^{-\g_5 \th_i -i\chi_i} \psi_{i\,0},\quad 
\bar\psi_i = \bar\psi_{i\,0} e^{-\g_5 \th_i +i\chi_i}.
\eeq
Then the bulk action becomes a free field one
\beq
S_{bulk} = \frac{1}{2\pi} \int_M d\t d\s \sum_{i=1}^2
\left[
\bar\psi_{i\,0} \gamma^\m \p_\m \psi_{i\,0} + \left(\frac{\pi}{g}
+1\right)\left(\p\th_i\right)^2 + \frac{\pi}{g}
\left(\p\chi_i\right)^2\right].
\eeq
The additional kinetic action for $\th_i$ is a manifestation of 
the $U(1)$ chiral anomaly: 
\beq
D[\psi]D[\bar\psi] = D[\psi_0] D[\bar\psi_0] \exp
\left[\frac{1}{4\pi} \int d\t d\s \sum_i (\p \th_i)^2 \right].
\eeq
Since the boundary mass term is not invariant under the $U(1)$ chiral gauge transformation, it transforms as 
\beq
\sum_i \bar\psi_i\psi_i = \sum_i \bar\psi_{i0} e^{-2\g_5 \th_i} \psi_{i0}.
\eeq
Note that scalar fields $\chi_i$ are free in the bulk and do not appear in the boundary action. Since the physical operators, being 
$U(1)_V$ gauge invariant, do not depend on $\chi_i$, we may drop
them. For the sake of convenience, we scale the scalar fields
\beq
\th_i \rightarrow \kappa \th_i, ~~~
\kappa = \sqrt{\frac{g}{2(\pi+g)}}.
\eeq
It brings us to 
\beq
S = \frac{1}{2\pi} \int_M d\t d\s \sum_{i=1}^2
\left[
\bar\psi_{i\,0} \gamma^\m \p_\m \psi_{i\,0} + \half\left(\p\th_i\right)^2 \right] + m \int_{\p M} d\s \sum_i \bar\psi_{i0} e^{-2\g_5 \kappa\th_i} \psi_{i0}.
\eeq
It is clear that in the limit of the critical point, $\kappa \rightarrow 0$, the interaction between the scalar fields
$\th_i$ and $\psi_0$ vanishes. Thus, $\th_i$, becoming free 
fields, can be dropped and the action reduces to that for free fermions with boundary masses at the critical point. 



The next step to construct the perturbative boundary state formulation is to find appropriate boundary conditions for 
$\psi^i_0$ and $\th_i$. We note that the boundary conditions for the fermion fields $\psi^i_0$
should coincide with those at the critical point. So the boundary conditions for them are the same as those in Eq.(\ref{simple}).
The boundary conditions for the fermion fields would remain intact as the Thirring interaction term is turned on, since the boundary conditions for the boson fields $\phi_i$ do not change. The boundary conditions for the fermion fields can be formally kept unchanged, if we require the boson fields to satisfy the following conditions
\beq \label{boundary2}
\th_1 |B_0\rangle = -\th_2 |B_0 \rangle,~~
\chi_1 |B_0\rangle = \chi_2 |B_0 \rangle .
\eeq
Since $\chi_i$ are completely decoupled from the physical degrees of freedom, the boundary conditions for $\chi_i$ are not important.
Note that Eq.(\ref{boundary2}) only fixes the boundary condition for $\frac{1}{\sqrt{2}}(\th_1 + \th_2)$. We may choose Neumann conditions for $\frac{1}{\sqrt{2}}(\th_1 - \th_2)$ for the sake of completeness. 
Once, the simple boundary state $|B_0\rangle$ is constructed,
the boundary state for the dissipated system $|B(m,\kappa)\rangle$ may be given as 
\beq
|B(m,\kappa) \rangle = \exp\left[m \int_{\p M} d\s \sum_i \bar\psi_{i} e^{-2\g_5 \kappa\th_i} \psi_{i} \right]|B_0\rangle
\eeq
where we drop the subscript $``\,{}_0\,"$ of the fermion fields for notational convenience. 

It is interesting to see that the scalar fields $\th_i$ appear
only through the boundary interaction and the physical operators
such as currents do not depend upon them.
Near the critical point where $|\kappa| < 1$, we may expand the boundary state $|B(m,\kappa)\rangle$ in $\kappa$ as follows
\beq
|B(m,\kappa)\rangle &=& \exp\left[m\int d\s \sum_i \bar\psi_i \psi_i 
-2m\kappa \int d\s \sum_i \bar\psi_i \g^5 \psi_i \th_i \right]|B(0,0)\rangle \nn\\
&=& \sum_n \frac{(-2m\kappa)^n}{n!}\left[\int d\s \sum_i \bar\psi_i \g^5 \psi_i \th_i \right]^n |B(m,0)\rangle 
\eeq
where $|B(m,0)\rangle$ corresponds to the boundary state at the critical point, of which explicit expression can be found in \cite{Hassel}.

As we expand $|B(m,\kappa)\rangle$ in $\kappa$, we encounter
a divergent term, proportional to the boundary mass term
 with a coefficient
\beq
&& ~ 2m^2 \kappa^2 \int d\s_1 \int d\s_2 \sum_i \langle \th_i(\s_1) \th_i(\s_2)\rangle \Bigl( \langle \bar \psi \g^5 \psi (\s_1) 
\bar \psi \g^5 \psi (\s_2) \rangle \nn\\
&& = 2m^2 \kappa^2 \int d\s_1 \int d\s_2 \, \ln \left\vert
1 - \frac{z_1}{z_2} \right\vert \left( \frac{1}{z_1 - z_2} 
\frac{1}{\barz_1 - \barz_2} \right)
\eeq
where $z_i = e^{i\s_i}$ and $\barz_i = e^{-i\s_i}$.
In order to regularize it we may deform the integration contours for 
$z_i$, introducing an infinitesimal parameter
$|\e| \ll 1 $ as follows
\beq
z_1 = e^{i\s_1}, ~~~ \barz_1 = \frac{1}{z}_1, ~~~
z_2 = e^{-\e} e^{i\s_2}, ~~~ \barz_2 = \frac{e^{-2\e}}{z_2}.
\eeq
Then we find,
\beq
\int d\s_1 \int d\s_2 \, \ln \left\vert
1 - \frac{z_1}{z_2} \right\vert \left( \frac{1}{z_1 - z_2} 
\frac{1}{\barz_1 - \barz_2} \right)
=  \frac{\pi^2}{2\e}. 
\eeq
This divergence can be taken care of by renormalization of the boundary mass. 
It yields 
\beq
m = m_0 \left[1 + \frac{\a-1}{2\a} \ln \frac{\Lambda^2}{\mu^2}\right]=
m_0 \left(\frac{\Lambda^2}{\mu^2}\right)^{\frac{(\a-1)}{2\a}}
\eeq
where $\ln \frac{\Lambda^2}{\mu^2} = \frac{m^2_0 \pi^2}{\e}$.
The result is in complete agreement with the previous work ref.\cite{fisher}: 
If $\a >1$, $m$ tends to grow and if $\a <1$, it scales to zero. The corection
to $m$ vanishes when $\a =1$. In fact, an explicit construction of the boundary state shows that perturbative corrections to $m$ vanish at all orders if $\a=1$.

\section{Inhomogeneous Rolling Tachyon}

In order to show that there exists a one parameter family of inequivalent solutions which describe the rolling of a generic open string tachyon away from its maximum, Sen \cite{sen022} discussed a D-$p$-brane of the bosonic string theory with one direction wrapped on a cricle of radius $R >1$. In addition to the usual tachyonic mode of mass$^2 = -1$, this system has a tachyonic mode of mass
\beq
R^{-2} -1 = -m^2,
\eeq
which comes from the first momentum mode of the standard tachyon along the circle direction. If we denote the Wick rotated time coordinate by
$X$ and the coordinate along the circle by $Y$, the conformal field theory associated with the rolling tachyon is obtained by perturbing the free conformal field theory by the boundary operator
\beq
g \int d\s \cos (m X) \cos \left(\frac{Y}{R}\right). 
\eeq
Thus, the rolling tachyon for the system with a compact spatial coordinate is described by the following action
\beq\label{compact}
S = \frac{1}{4\pi \ap} \int_M d\t d\s \left(\p X \p X + \p Y \p Y \right) + g \int_{\p M} d\s \cos (m X) \cos \left(\frac{Y}{R}\right).
\eeq
Since the boundary perturbation depends not only on $X$ but also on the spatial coordinate $Y$, we expect that the tachyon condensation may be inhomogeneous.

It may be convenient to rewrite the action 
in terms of scalar fields, $\phi_i$, $i=1,2$ defined as 
\beq
\phi_1 &=& mX + \frac{Y}{R} = \cos \th\, X + \sin \th\, Y, \nn\\
\phi_2 &=& mX - \frac{Y}{R} = \cos \th\, X - \sin \th\, Y
\eeq
where 
\beq
m = \cos \th, ~~~ \frac{1}{R} = \sin \th.
\eeq
Rewriting the action in terms of $\phi_i$, we have 
\beq
S = \frac{1}{4\pi} \int_M d\t d\s \left[\p \phi_a \p \phi_a + g^{ab} \p \phi_a \p \phi_b \right] + \frac{g}{4}\int_{\p M} d\s \sum_a \left(
e^{i\phi_a} + e^{-i\phi_a} \right)
\eeq
where
\beq
g^{11} = g^{22} = \left(\frac{1}{\ap \sin^2 (2\th)} -1 \right),~~~
g^{12} = g^{21} = - \frac{1}{\ap} \frac{\cos 2\th}{\sin^2(2\th)}.
\eeq
It is easy to see that when $g^{ab} = 0$, {\it i.e.}, 
\beq
\ap =1,~~~ \th = \frac{\pi}{4},
\eeq
the action reduces to that for a critical theory with a sum of two commuting exactly marginal perturbation
\beq
S = \frac{1}{4\pi} \int_M d\t d\s\,\p \phi_a \p \phi_a + \frac{g}{4}\int_{\p M} d\s \sum_a \left(
e^{i\phi_a} + e^{-i\phi_a} \right).
\eeq
The critical theory becomes a two flavor
$SU(2) \times SU(2)$ free fermion theory with a boundary mass if fermionized. The time evolution of the rolling tachyon at this critical point has been discussed by Sen \cite{sen022}.

As in the homogeneous tachyon condensation, it would be convenient to employ the fermionization technique to analyze the critical behaviour of the system. It begins with diagonalizing the bulk action. The bulk action can be diagonalized by a similarity transformation
for the fields $\phi_a$
\beq
\phi_a = \sum_b M_{ab} \phi^\prime_b, ~~~
M = M^\dagger = M^{-1} = \frac{1}{\sqrt{2}} 
\left(\begin{array}{rr}
1 & 1 \\
1 & -1 \end{array} \right),\qquad  M^t M = I.
\eeq
By the similarity transformation the Lagrangian is diagonalized as\beq
L_0 = \frac{1}{4\pi} \sum_a \p \phi^\prime_a \p \phi^\prime_a + \frac{1}{4\pi} \sum_{a} \lambda_a \p \phi^\prime_a \p \phi^\prime_a 
\eeq
where
\beq
\lambda_1 = \frac{1}{2\ap \cos^2 \th} -1, ~~~
\lambda_2 = \frac{1}{2\ap \sin^2 \th} -1.
\eeq
Then we may introduce auxiliary boson fields $\varphi^\prime_a$, $i =,1,2$ for which the action has the same form as 
that for $\phi^\prime_a$
\beq
L_0 = \frac{1}{4\pi} \sum_a \left(1 + \lambda_a \right)\p \phi^\prime_a \p \phi^\prime_a +  \frac{1}{4\pi} \sum_a \left(1 + \lambda_a \right)\p \varphi^\prime_a \p \varphi^\prime_a.
\eeq
Since the bulk action for $\varphi^\prime_a$ is the free field one, $\varphi^\prime_a$ would decouple from $\phi^\prime_a$ if the Dirichlet boundary condition is chosen for $\varphi^\prime_a$. It also follows from that $\varphi_a$ fields would decouple from the physical fields $\phi_a$ since fields $\phi_a$ and
$\varphi_a$ are related to fields $\phi^\prime_a$ and 
$\varphi^\prime_a$ respectively by a linear transformation. 

Defining $\Phi^\prime_{a i}$ as follows
\beq
\Phi^\prime_{a1} = \frac{1}{\sqrt{2}}\left(\phi^\prime_a + \varphi^\prime_a\right),~~~
\Phi^\prime_{a2} = \frac{1}{\sqrt{2}}\left(\phi^\prime_a - \varphi^\prime_a \right), ~~~ a =1, 2,
\eeq
we may rewrite the bulk Lagrangian as 
\beq
L_0 = \frac{1}{4\pi} \sum_{a,i} \p \Phi^\prime_{a i} \p \Phi^\prime_{a i} + \frac{1}{4\pi} \sum_{a,i}\lambda_a \p \Phi^\prime_{a i} \p \Phi^\prime_{b i} 
\eeq
Taking the similarity transformation again
\beq
\Phi_{ai} = M^{-1}_{ab} \Phi^\prime_{b i}, ~~~ \varphi_{a i} = M^{-1}_{ab} 
\varphi^\prime_{b i} ,
\eeq
we have
\beq
L_0= \frac{1}{4\pi} \sum_a \sum_i \p \Phi_{a i} \p \Phi_{a i} + 
\frac{1}{4\pi} \sum_{a,b} \sum_i g^{ab} \p \Phi_{a i} \p \Phi_{b i} ~~. \nn
\eeq

Now we are in the position to fermonize the system. Introducing the fermion fields defined as 
\beq
\psi_{a1L} &=& \zeta_{a1L} :e^{-i\sqrt{2} \Phi_{a1L}}:, ~~~
\psi_{a2L} = \zeta_{a2L} :e^{i\sqrt{2} \Phi_{a2L}}: \\
\psi_{a1R} &=& \zeta_{a1R} :e^{i\sqrt{2} \Phi_{a1R}}:, ~~~
\psi_{a2R} = \zeta_{a2R} :e^{-i\sqrt{2} \Phi_{a2R}}:, \nn
\eeq
we find that the bulk action can be written as 
\beq
S_M = \frac{1}{2\pi} \int_M d\t d\s \left(\sum_{a,i} \bar\psi_{ai} \g^\m \p_\m \psi_{ai} + \sum_{a,b}\sum_i\frac{g^{ab}}{4} j^\m{}_{ai} j_\m{}_{bi} \right).
\eeq
Since we choose the Neumann condition as the boundary condition for 
$\phi_a$ and the Dirichlet condition for $\varphi_a$, the boundary conditions for the fields $\Phi_{ai}$ are given as 
\beq
\Phi_{a1L}(0,\s)|B_0\rangle  = \Phi_{a2R}(0,\s)|B_0\rangle, ~~~
\Phi_{a2L}(0,\s)|B_0\rangle  = \Phi_{a1R}(0,\s)|B_0\rangle .
\eeq
It follows that with the chosen boundary condition the boundary potential may be written as 
\beq
L_B = g \sum_a \left(e^{i\phi_a} + e^{-i\phi_a} \right) = 
g \sum_a \left(e^{i\sqrt{2} \Phi_a} + e^{-i\sqrt{2} \Phi_a} \right).
\eeq
This boundary potential term is equivalent to a boundary mass term in the fermion theory. Hence, the fermionization finally brings us to the following generalized Thirring model with a boundary mass 
\beq\label{fermionaction}
S = \frac{1}{2\pi} \int_M d\t d\s \left(\sum_{a,i} \bar\psi_{ai} \g^\m \p_\m \psi_{ai} + \sum_{a,b}\sum_i\frac{g^{ab}}{4} j^\m{}_{ai} j_\m{}_{bi} \right) + \frac{g}{2} \int_{\p M} d\s \sum_{a,i} \bar\psi_{ai} \psi_{ai} ~ .
\eeq

Appling the renormalization group (RG) analysis, developed in the section 3, to this generalized Thirring model, we find that the RG flow of $g$ does not depend on $R$
\beq
g = g_0 \left[1 + \frac{1-\ap}{2} \ln \frac{\Lambda^2}{\mu^2} \right]=
g_0 \left[\frac{\Lambda^2}{\mu^2}\right]^{\frac{1-\ap}{2}}. 
\eeq
The inhomogeneous rolling tachyon corresponds to the case where $\ap =1$. However, a more detailed RG analysis reveals that this is not the end of story: The generalized Thirring model and consequently, the
inhomogeneous rolling tachyon turns out to have non-trivial phase diagrams. The action Eq.(\ref{compact}) or its fermionized one Eq.(\ref{fermionaction}) is in fact equivalent to the 
spin-dependent Tomonaga-Luttinger model with a scattering potential at the origin, which describes quantum transport through a single barrier in a one-dimensional interacting electron system, discussed by Furusaki and Nagaosa \cite{furusaki}. When we deals with the perturbation theory, we must also consider the renormalization of the descendents of the boundary mass operators such as 
$g_+ e^{i\sqrt{2}(\Phi_1 + \Phi_2)}$ and $g_-e^{i\sqrt{2}(\Phi_1-\Phi_2)}$.
Applying the RG analysis to these descendent operators, which arise in the second-order perturbation, we obtain the RG equations for $g_\pm$,
\beq
g_+ = g_{+0} \left[\frac{\Lambda^2}{\m^2}\right]^{1-4\ap\left(1-R^{-2}\right)}, ~~~
g_- = g_{-0} \left[\frac{\Lambda^2}{\m^2}\right]^{1-4\ap R^{-2}}.
\eeq
Thus, the RG equations for the boundary mass operator and its descendents yield that the phase diagram divides into four regions as shown in Fig. 1 and the phase boundaries are set by 
\beq
\a = \frac{1}{\ap}=1,~~~ \frac{1}{R^2} = \frac{\a}{4},~~~
\frac{1}{R^2} = 1 - \frac{\a}{4}.   
\eeq
(One may obtain the same phase diagram by transcribing the result of Furusaki and Nagaosa \cite{furusaki} into the string theory.) 

\begin{figure}[htbp]
   \begin {center}
\setlength{\unitlength}{2mm}
\ifx\plotpoint\undefined\newsavebox{\plotpoint}\fi
\begin{picture}(70,50)(0,0)
\put(5,5){\vector(1,0){55}}
\put(5,5){\vector(0,1){45}}
\put(65,5){\makebox(0,0)[tl]{\large$\a=1/\ap$}}
\put(5,50){\makebox(0,0)[br]{\large${R^{-2}}$}}
\put(5,45){\line(1,0){55}}
\put(45,5){\line(0,1){40}}
\put(5,5){\line(4,1){40}}
\put(5,45){\line(4,-1){40}}
\put(45,3){\makebox(0,0)[t]{\large 1}}
\put(3,45){\makebox(0,0)[r]{\large 1}}
\put(3,3){\makebox(0,0)[r]{\large $0$}}
\put(57,25){\makebox(2,2)[c]{\Large IV}}
\put(37,40){\makebox(2,2)[c]{\Large I}}
\put(37,7){\makebox(2,2)[c]{\Large III}}
\put(22,25){\makebox(2,2)[c]{\Large II}}
\put(45,5){\circle*{1}}
\put(45,25){\circle*{1}}
\end{picture}
   \end {center}
   \caption {\label{phase1} Phase Diagram for The Generalized Thirring Model}
\end{figure}
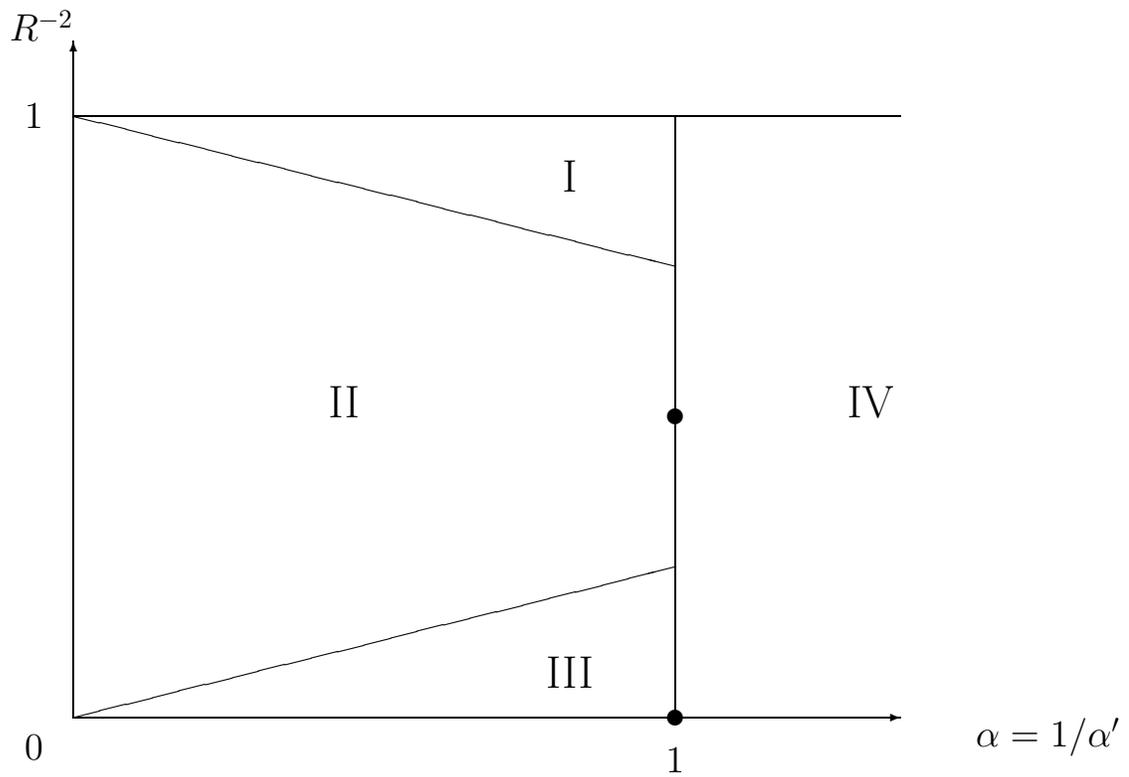

In the regions I, II, III and IV the RG flow leads the system to fixed points, which correspond to the boundary states $|D,N\rangle$, $|N,N\rangle$, $|N,D\rangle$ and $|D,D\rangle$ respectively. Here $N$ and $D$ denote Neumann and Dirichlet boundary conditions and the first and second label is the condition for the $X$ and $Y$ bosons, respectively. Consequently if we begin with a $Dp$-brane with the boundary interaction, due to the tachyon condensation, the RG flow 
drives the system into $Sp$-brane, $Dp$-brane, $D(p-1)$-brane, and
$S(p-1)$-brane in the regions I, II, III, and IV respectively.  

If we wish to study the inhomogeneous rolling tachyon, we should pay
attention to the line, $\a =1$ on the phase diagram Fig. 1. From the 
phase diagram Fig. 1, it is clear that the inhomogeneous rolling tachyon model has three different phases as depicted in Fig. 2: 
A) $1 < R < \frac{2}{\sqrt{3}}$, B) $\frac{2}{\sqrt{3}} < R < 2$,
C) $R > 2$. The point where $R = \sqrt{2}$ corresponds to the 
critical point where the model reduces to the $SU(2) \times SU(2)$ free fermion theory with boundary masses discussed by Sen 
\cite{sen022}.

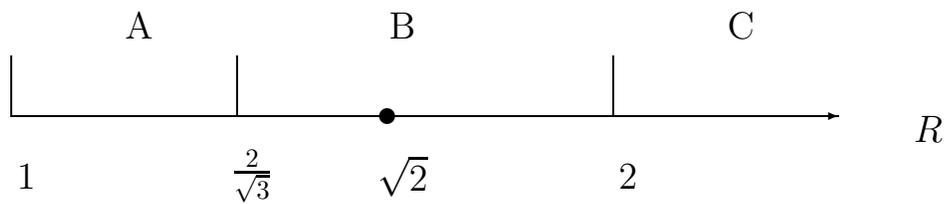
\begin{figure}[htbp]
   \begin {center}
\setlength{\unitlength}{2mm}
\ifx\plotpoint\undefined\newsavebox{\plotpoint}\fi
\begin{picture}(70,20)(0,0)
\put(5,5){\vector(1,0){55}}
\put(65,5){\makebox(0,0)[tl]{\large $R$}}
\put(5,0){\makebox(2,2)[c]{\large $1$}}
\put(20,0){\makebox(2,2)[c]{\large $\frac{2}{\sqrt{3}}$}}
\put(30,0){\makebox(2,2)[c]{\large $\sqrt{2}$}}
\put(45,0){\makebox(2,2)[c]{\large $2$}}
\put(12.5,10){\makebox(2,2)[c]{\large A}}
\put(30,10){\makebox(2,2)[c]{\large B}}
\put(52.5,10){\makebox(2,2)[c]{\large C}}
\put(5,5){\line(0,1){4}}
\put(20,5){\line(0,1){4}}
\put(45,5){\line(0,1){4}}
\put(30,5){\circle*{1}}
\end{picture}
   \end {center}
   \caption {\label{phase2} Phase Diagram for Inhomogeneous Rolling Tachyon}
\end{figure}

If we take the RG effects into account, we 
may have to choose a different action for each region to describe the
inhomogeneous rolling tachyon in a way consistent with the perturbation theory. In the region A the boundary operator 
$e^{i\sqrt{2}(\Phi_1 + \Phi_2)}$ is a relevant operator while
$e^{i\sqrt{2}(\Phi_1-\Phi_2)}$ is an irrelevant one. The RG effects drive 
the boundary condition for $X$ to be Dirichlet. If the Dirichlet condition is once chosen for $X$, the action is reduced as 
\beq
S = \frac{1}{4\pi} \int_M d\t d\s \left(\p X \p X + \p Y \p Y \right) + \frac{g}{2}\int_{\p M}
d \s \left(e^{i\sqrt{1-R^{-2}}X} + e^{-i\sqrt{1-R^{-2}}X}\right).
\eeq
Note that the boundary mass operator becomes irrelevant in the region A. Hence, they can be ignored. It follows that the action for the rolling tachyon, compatible with the perturbation theory, 
is given as a sum of two free boson actions (or a free fermion action equivalently) in the region A
\beq\label{freeboson}
S = \frac{1}{4\pi} \int_M d\t d\s \left(\p X \p X + \p Y \p Y \right)
\eeq
where the boundary conditions for $X$ and $Y$ are Dirichlet and Neumann respectively. Since the boundary conditon for $X$ is chosen to be Dirichlet, the tachyon is not rolling and the theory describes a static $Sp$-brane.
In the region B both boundary operators
$e^{i\sqrt{2}(\Phi_1 + \Phi_2)}$ and $e^{i\sqrt{2}(\Phi_1-\Phi_2)}$ are irrelvant, but the boundary mass operators $\bar\psi_{ai} \psi_{ai}$ are marginal. Thus, the inhomogeneous tachyon rolls and 
the rolling tachyon is described by the Thirring action
Eq.(\ref{fermionaction}) with $g^{ab}$ given as follows
\beq
g^{11} = g^{22} = \left(\frac{1}{\sin^2 (2\th)} -1 \right),~~~
g^{12} = g^{21} = - \frac{\cos 2\th}{\sin^2(2\th)}
\eeq
in the region B.
In the region C the boundary operator 
$e^{i\sqrt{2}(\Phi_1 + \Phi_2)}$ is an irrelevant operator while
$e^{i\sqrt{2}(\Phi_1-\Phi_2)}$ is a relevant one. The RG effects drive the boundary condition for $Y$ to be Dirichlet. Then if the Dirichlet condition
is chosen as the boundary condition for $Y$, the action may be written as
\beq
S = \frac{1}{4\pi} \int_M d\t d\s \left(\p X \p X + \p Y \p Y \right) + \frac{g}{2}\int_{\p M}
d \s \left(e^{i\frac{Y}{R}} + e^{-i\frac{Y}{R}}\right).
\eeq
The boundary operator, $e^{i\frac{Y}{R}} + e^{-i\frac{Y}{R}}$, is irrelevant in the region C. Thus, the action, which is compatible with the perturbation theory, reduces to a sum of two free boson actions Eq.(\ref{freeboson}) as in the region A. However,
the boundary conditions differ from those in the region A: 
The boundary conditions for $X$ and $Y$ are Neumann and Dirichlet respectively. The action depicts a static $D(p-1)$-brane instead of a time
dependent evolution of the system. Based on the RG analysis, we may conclude that the inhomogeneous tachyon is rolling only when $\frac{2}{\sqrt{3}} < R < 2$. In the region C one may further notice that the boundary mass operators become marginal as $R \rightarrow \infty$ (with the Dirichlet boundary condition for $Y$) and the tachyon condensation may depend on time. This special point corresponds to the homogeneous rolling tachyon discussed in the sections 2 and 3.                      

\section{Conclusions}

A few remarks are in order to conclude this paper. We first discuss the dissipative quantum system in one dimension and its relation to the homogeneous rolling tachyon, employing the boundary state formulation of string theory and the Thirring model in two dimensions. The framework presented here has some advantages over the previous ones on the dissipative quantum systems. We need to deal with a local boundary interaction only and can take an advantage of the string theory techniques to explore various aspects of the system. Especially, the boundary state
for the system can be explicitly constructed.
Since the system is described in terms of the free fields, all the physical quantities are exactly calculable at any given order. 

The boundary state formulation of the Thirring model with a boundary 
mass is also found to be useful to study the inhomogeneous tachyon 
\cite{sen022} and the single-barrier problem in a one-dimensional interacting electron system \cite{furusaki}. We show then if fermionized, both models are described by a generalized Thirring action with boundary masses 
\beq \label{general}
S = \frac{1}{2\pi} \int_M d\t d\s \left(\sum_{a,i =1}^2 \bar\psi_{ai} \g^\m \p_\m \psi_{ai} + \sum_{a,b=1}^2\sum_{i=1}^2\frac{g^{ab}}{4} j^\m{}_{ai} j_\m{}_{bi} \right) + \frac{g}{2} \int_{\p M} d\s \sum_{a,i=1}^2 \bar\psi_{ai} \psi_{ai} .
\eeq
The inhomogeneous rolling tachyon is a special case where $\ap = 1/\a =1$. From the RG analysis of the boundary mass and its descendent operators in the perturbation theory, we point out that the generalized Thirring model has a non-trivial phase diagram which divides into four region, depending on the 
string tension and the radius of the compact circle.
The RG fixed point for each region corresponds to $Sp$-brane, $Dp$-brane, $D(p-1)$-brane, and $S(p-1)$-brane respectively. Concurrently, the inhomogeneous rolling tachyon model also has a non-trivial phase diagram, which divides into
three different regions, depending on the radius of the compact circle. 
Since in the region where $1 < R < \frac{2}{\sqrt{3}}$ and 
$R > 2$, the boundary mass operators and its descendents have trivial fixed
points, corresponding to the Dirichlet state or the Neumann state,
the RG flow drives the action to free field one which describes a static $Sp$-brane or a $D(p-1)$-brane. The inhomogeneous tachyon rolls only when
$\frac{2}{\sqrt{3}} < R < 2$. 

There are several directions along which this work can be extended. We construct a fermion Thirring model, which is more suitable to analyze the dynamics of the inhomogeneous rolling tachyon than the bosonic model. The constructed Thirring model would be also useful to study the single-barrier problem in condensed matter physics near the critical points. One of advantages of the fermion formulation is that
it is easier to construct boundary states, which are extremely convenient to discuss various dynamical aspects of the theories. 
We save more detailed analyses of the inhomogeneous tachyon condensation and the single-barrier problem based on the boundary state formulation for a future work \cite{cholee2007}. We may also employ the constructed Thirring to evaluate the closed string emission from the decaying $Dp$-brane through the inhomogeneous tachyon condensation.   

The string theory action
Eq.(\ref{sine1}) or Eq.(\ref{sine2}) for the dissipative system is also known as the boundary sine-Gordon model, which has been used to study the quantum impurity problems and the edge state
tunnelling in the fractional quantum Hall effect. See ref.\cite{saleur}
and references there in for discussions in this direction. The quantum integrability plays an important role in the studies along 
this direction. It may be interesting to explore the relationship between the boundary state formulation given in this paper and the analysis based on the quantum integrability. The quantum integrability
of the more general boundary sine-Gordon models, corresponding to the 
Thirring models for the inhomogeneous tachyon condensation and the 
single-barrier problem also deserves to be an interesting subject to be studied.

\vskip 2cm

\noindent {\bf Acknowledgement:} The author thanks Gordon Semenoff, Philip Stamp and Jin-Ho Cho for informative discussions. Part of this work was done during the author's visit to KIAS (Korea) and PITP (Canada).


\begin{thebibliography}{99}

\bibitem{schmid}
A.~Schmid, 
``Diffusion and localization in a dissipative quantum system,"
Phys. Rev. Lett. {\bf 51}, 1506 (1983).


\bibitem{Guinea}
  F.~Guinea, V.~Hakim and A.~Muramatsu,
  ``Diffusion and localization of a particle in a periodic
  potential coupled to a dissipative environment,''
  Phys. Rev. Lett. {\bf 54}, 263 (1985).

\bibitem{fisher}
M.P.A.~Fisher and W.~Zwerger, 
``Quantum Brownian motion in a periodic potential,'' 
Phys. Rev. {\bf B32}, 6190 (1985).

\bibitem{furusaki}
A. Furusaki and N. Nagaosa,
``Single-baarrier problem and Anderson localization in a one-dimensional interacting electron system,"'
Phys. Rev. {\bf B47}, 4631 (1993).


\bibitem{Callan:1989mm}
  C.~G.~Callan and L.~Thorlacius,
  ``Open String Theory As Dissipative Quantum Mechanics,''
  Nucl.\ Phys.\ B {\bf 329}, 117 (1990).


\bibitem{larkin}
L.I.~Glazman and A.I.~Larkin, 
``New quantum phases in a one-dimensional junction array,''   
Phys.Rev.Lett. {\bf 79}, 3736 (1997).
 [arXiv:cond-mat/9809118].



\bibitem{fazio}
 R. Fazio and H. van der Zant, 
 ''Quantum Phase transitions and Vortex dynamics in Superconducting networks'', 
 Phys. Rep. {\bf 355}, 235 (2001)


\bibitem{sodano}
D.~Giuliano and P.~Sodano, 
``Effective boundary field theory for a Josephson junction chain with a weak link'', 
Nucl. Phys. B711, 480-504, (2005), [arXiv:cond-mat/0501378].



\bibitem{Callan:1991da}
  C.~G.~Callan and D.~Freed,
``Phase diagram of the dissipative Hofstadter model,''
  Nucl.\ Phys.\ B {\bf 374}, 543 (1992), [arXiv:hep-th/9110046].

\bibitem{Callan:1992vy}
  C.~G.~Callan, A.~G.~Felce and D.~E.~Freed,
 ``Critical theories of the dissipative Hofstadter model,''
  Nucl.\ Phys.\ B {\bf 392}, 551 (1993), [arXiv:hep-th/9202085].


\bibitem{Affleck:1990iv}
  I.~Affleck and A.~W.~W.~Ludwig, ``Critical Theory Of Overscreened Kondo Fixed Points,''
  Nucl.\ Phys.\ B {\bf 360}, 641 (1991).

\bibitem{Affleck:1990by}
  I.~Affleck and A.~W.~W.~Ludwig,
 ``The Kondo Effect, Conformal Field Theory And Fusion Rules,''
  Nucl.\ Phys.\ B {\bf 352}, 849 (1991).


\bibitem{Kane:1992}
  C.~L.~Kane and M.~P.~A.~Fisher,
  ``Transmission through barriers and resonant tunneling in an interacting one-dimensional electron gas,''
Phys. Rev. {\bf B46}, 15233 (1992).

 \bibitem{kane2}
 C.L.~Kane and M.P.A.~Fisher, 
 ``Contacts and Edge State Equilibration in the Fractional
 Quantum Hall Effect''  
[arXiv:cond-mat/9506116].




\bibitem{Oshikawa:2005fh}
 C.~Chamon,  M.~Oshikawa and I.~Affleck,
  ``Junctions of three quantum wires,''
  Phys.\ Rev.\ Lett.\  {\bf 91}, 206403 (2003), [arXiv:cond-mat/0509675].

\bibitem{Witten:1992qy}
  E.~Witten,
  ``On background independent open string field theory,''
  Phys.\ Rev.\ D {\bf 46}, 5467 (1992), [arXiv:hep-th/9208027].

\bibitem{Shatashvili:1993kk}
  S.~L.~Shatashvili,
 ``Comment on the background independent open string theory,''
  Phys.\ Lett.\ B {\bf 311}, 83 (1993), [arXiv:hep-th/9303143].

\bibitem{Shatashvili:1993ps}
  S.~L.~Shatashvili,
  ``On the problems with background independence in string theory,''
  Alg.\ Anal.\  {\bf 6}, 215 (1994), [arXiv:hep-th/9311177].

\bibitem{Tlee:01nc}
 T.~Lee,
 ``Tachyon condensation, boundary state and noncommutative solitons,"
 Phys.\ Rev.\ D {\bf 64}, 106004 (2001), [arXiv:hep-th/0105115]. 
 
\bibitem{Tlee:01os}
 T.~Lee,
 ``Tachyon condensation and open string field theory,"
 Phys.\ Lett.\ B {\bf 520}, 385 (2001), [arXiv:hep-th/0105264].
 
\bibitem{Tlee:01one}
 T.~Lee,
``Boundary string field theory at one-loop,"
 J.\ Korean\ Phys.\ Soc.\ {\bf 42}, 34 (2003), [arXiv:hep-th/0109032].
 
\bibitem{Sen:2002nu}
  A.~Sen, ``Rolling tachyon,''
  JHEP {\bf 0204}, 048 (2002), [arXiv:hep-th/0203211].
  
\bibitem{senreview} See for a review on the rolling 
tachyon:
A.~Sen, ``Tachyon dynamics in open string theory", Int. J. Mod. Phys. {\bf A20} 
(2005) 5513, [arXiv:hep-th/0410103].
 
\bibitem{lambert}  N. Lambert, H. Liu and J. Maldacena, ``Closed strings from
decaying D-branes", [arXiv:hep-th/0303139].

 \bibitem{sen022}  A.~Sen, ``Time evolution in open string
theory,''  JHEP {\bf 0210}, 003 (2002), [arXiv:hep-th/0207105].

\bibitem{mukhopa}  P.~Mukhopadhyay and A.~Sen, ``Decay of
unstable D-branes with electric field,''  JHEP {\bf 0211}, 047
(2002), [arXiv:hep-th/0208142].

\bibitem{sen023}   A.~Sen, ``Time and tachyon", [arXiv:hep-th/0209122].

\bibitem{sen0305} A. Sen, ``Open and closed strings from unstable 
D-branes, Phys. Rev. {\bf D68} (2003) 106003, [arXiv:hep-th/0305011].

\bibitem{sen031}  A.~Sen, ``Dirac-Born-Infeld Action on the
Tachyon Kink and Vortex", [arXiv:hep-th/0303057].

\bibitem{sen0306} A. Sen, ``Open-Closed duality at tree level",
Phys. Rev. Lett. {\bf 91} (2003) 181601, [arXiv:hep-th/0306137]

\bibitem{sen0308} A. Sen, ``Open-Closed duality: Lessons from 
matrix model", Mod. Phys. Lett. {\bf A19} (2004) 841, [arXiv:hep-th/0308068]

\bibitem{sen04}  A. Sen, ``Rolling tachyon boundary
state, conserved charges and two dimensional string theory", JHEP
{\bf 0405}, 076 (2004), [arXiv:hep-th/0402157]

\bibitem{larsen02}  F.~Larsen, A.~Naqvi and S.~Terashima,
``Rolling tachyons and decaying branes", JHEP {\bf 0302} (2003) 039,
[arXiv:hep-th/0212248].

\bibitem{gibbons} G. W. Gibbons, K. Hori and P. Yi,
``String fluid  from unstable D-branes", Nucl. Phys. {\bf B596} (2001) 136, [arXiv:hep-th/0009061].

\bibitem{okuda} T. Okuda and S. Sugimoto, ``Coupling of rolling 
tachyon to closed strings", Nucl. Phys. {\bf B647} (2002) 101, 
[arXiv:hep-th/0208196]

\bibitem{kim} C. Kim, H. B. Kim and Y. Kim, ``Rolling tachyons in 
string cosmology", Phys. Lett. {\bf B552} (2003) 111, [arXiv:hep-th/0210101]

\bibitem{hlee} H. Lee and W. S. l'Yi, ``Time evolution of rolling 
tachyons for a brane-anti-brane pair", J. Korean Phys. Soc. {\bf 43} 
(2003) 676, [arXiv:hep-th/0210221]

\bibitem{rey} S.~J.~Rey and
S.~Sugimoto,  ``Rolling tachyon with electric and magnetic fields:
T-duality approach", [arXiv:hep-th/0301049].

\bibitem{rey2} S.~J.~Rey and S.~Sugimoto, ``Rolling of modulated tachyon with gauge flux and emergent fundamental string", 
Phys. Rev. {\bf D68} (2003) 026003, [arXiv:hep-th/0303133] 


\bibitem{taka}  T. Takayanagi and N. Toumbas, ``A matrix model
dual of type 0B  string theory in two dimensions", [arXiv:hep-th/0307083].

\bibitem{doug}  M. R. Douglas, I. R. Klebanov, D. Kutasov, J.
Maldacena, E.  Martinec, and N. Seiberg, ``A new hat for the $c=1$
matric model", [arXiv:hep-th/0307195].

\bibitem{arefeva} I. Ya. Aref'eva, L. V. Joukovshaya and A. S. 
Koshelev, ``Time evolution in superstring field theory on non-BPS 
brane, JHEP {\bf 0309} (2003) 012, [arXiv:hep-th/0301137]


\bibitem{demasure} Y. Demasure and R. A. Janik, ``Backreaction and 
the rolling tachyon", Phys. Lett. {\bf B578} (2004) 195, [arXiv:hep-th/0305191]

\bibitem{gutperle03}  M.~Gutperle and A.~Strominger,
``Timelike boundary Liouville theory,''  [arXiv:hep-th/0301038].

\bibitem{larsen} F. Larsen, A. Naqvi and S. Terashima, ``Rolling 
tachyons and decaying branes", JHEP {\bf 0302} (2003) 039, 
[arXiv:hep-th/0212248]


\bibitem{constable} N. R. Constable and F. Larsen, ``The rolling 
tachyon as a matrix model", JHEP {\bf 0306} (2003) 017, [arXiv:hep-th/0305177]

\bibitem{schomerus} V. Schomerus, ``Rolling tachyons from 
Liouville theory", JHEP {\bf 0311} (2003) 043, [arXiv:hep-th/0306026]


\bibitem{nagami}
K. Nagami, ``Rolling tachyon with electromagnetic field in linear 
dilaton background", Phys. Lett. {\bf B591} (2004) 187, [arXiv:hep-th/0312149]

\bibitem{foto} A. Fotopoulos and A. A. Tseytlin, ``On open 
superstring partition function in inhomogeneous rolling tachyon
background", JHEP {\bf 0312} (2003) 025, [arXiv:hep-th/0310253]

\bibitem{coletti} E. Coletti, I. Sigalov, W. Taylor,
``Taming the Tachyon in Cubic String Field Theory",
JHEP {\bf 0508} (2005) 104, [arXiv:hep-th/0505031] 

\bibitem{forini} V. Forini, G. Grignani, G. Nardelli,
``A new rolling tachyon solution of cubic string field theory"
[hep-th/0502151]; ``A solution to the 4-tachyon off-shell amplitude in 
cubic string field theory", JHEP {\bf 0604} (2006) 053,
[arXiv:hep-th/0603206] 
  
\bibitem{Tlee:06}
 T.~Lee,
``The final fate of the rolling tachyon,"
 JHEP {\bf 0611}, 056 (2006),
 [arXiv:hep-th/0606236]
 
\bibitem{gianluca}
G. Calcagni and G. Nardelli, 
``Tachyon solutions in boundary and cubic string field theory,"
[arXiv:hep-th/0708.0366]

\bibitem{Tlee:2005ge}
  T.~Lee and G.~W.~Semenoff,
  ``Fermion representation of the rolling tachyon boundary conformal
 field theory,''
 JHEP {\bf 0505}, 072 (2005),
[arXiv:hep-th/0502236].

\bibitem{Hassel}
 M.~Hasselfield, T.~Lee, G.~W.~Semenoff and P.~C.~E.~ Stamp,
 ``Critical boundary sine-Gordon revisited,"
 Ann.\ Phys.\ {\bf 321}, 2849 (2006), [arXiv:hep-th/].
  
\bibitem{Thirring}
 W.~Thirring,
 ``A Soluble relativistic eld theory,"
 Ann.\ Phys.\ {\bf 3}, 91 (1958).
 
 \bibitem{Glaser}
 V. Glaser,
 ``An explicit solution of the Thirring model,"
 Nuovo Cim.\ {\bf 9}, 990 (1958).
 
 \bibitem{Johnson}
 K.~Johnson,
 ``Solution of the equations for the Green's functions of a two- dimensional relativistic field theory,"
 Nuovo Cim.\ {\bf 20}, 773 (1961).
 
 \bibitem{Hagen}
 C.~R.~Hagen,
 ``New solutions of the Thirring model,"
 Nuovo Cim.\ {\bf 51B}, 169 (1967).
 
 \bibitem{Klaiber}
 B. Klaiber, in ``Lectures in Theoretical Physics," edited by A.~Barut and W.~Brittin, Gordon and Breach, New York, 1968, p. 141-176. 

\bibitem{Schwinger}
 J.~S.~Schwinger,
``Gauge invariance and mass,"
 Phys.~Rev.~{\bf 125}, 397 (1962).
 
\bibitem{Coleman}
 S.~Coleman,
``Quantum sine-Gordon equation as the massive Thirring model,"
 Phys.~Rev.~ D {\bf 11}, 2088 (1975).
 
\bibitem{ji}
S. Ji, J.-Y. Koo and T. Lee,
``Dissipative Hofstadter model at the magic points and critical boundary Sine-Gordon model,"
J.\ Korean\ Phys.\ Soc. {\bf 50}, S54 (2007).
  
\bibitem{caldeira83ann} 
A. O. Caldeira and A. J. Leggett, 
``Quantum tunneling in a dissipative system,"
Ann. Phys. {\bf 149}, 374 (1983). 

\bibitem{caldeira83phy}
A. O. Caldeira and A. J. Leggett,
``Path integral approach to quantum Brownian motion,"
Physica {\bf 121A}, 587 (1983).

\bibitem{Polchinski:1994my}
J.~Polchinski and L.~Thorlacius,
``Free fermion representation of a boundary conformal field theory,''
Phys.\ Rev.\ D {\bf 50}, R622 (1994), [arXiv:hep-th/9404008].

\bibitem{saleur}
H. Saleur, 
`` Lectures on non-perturbative field theory and quantum impurity problems," [arXiv:cond-mat/9812110]; 
Lectures on non-perturbative field theory and quantum impurity problems: Part II," [arXiv:cond-mat/0007309].

\bibitem{cholee2007}
J.-H. Cho and T. Lee, in preparation (2007).


\end{thebibliography}
\end{document}